\documentclass[a4paper,twocolumn,10pt]{revtex4}
\usepackage{amssymb,amsmath,amsthm}
\usepackage{centernot}

\theoremstyle{plain}

\theoremstyle{definition}

\newcommand*{\eps}{\varepsilon}

\newcommand*{\Prob}{\mathrm{Prob}}

\begin{document}
\title{Reply to recent scepticism about the foundations of quantum
  cryptography}

\author{Renato Renner}

\affiliation{Institute for Theoretical Physics; ETH Zurich;
  Switzerland}

\date{September 11, 2012}

\begin{abstract}
  In a series of recent papers, Hirota and Yuen claim to have
  identified a fundamental flaw in the theory underlying quantum
  cryptography, which would invalidate existing security proofs. In
  this short note, we sketch their argument and show that their
  conclusion is unjustified|it originates from a confusion between
  necessary and sufficient criteria for secrecy.
\end{abstract}

\maketitle

The purpose of this note is to refute a critique by
Hirota~\cite{Hirota12} and
Yuen~\cite{Yuen12b,Yuen12a,Yuen11c,Yuen11b,Yuen11a,Yuen10,Yuen09b,Yuen09a}
concerning a basic criterion for secrecy~\cite{BHLMO05,RenKoe05},
which is widely used in quantum cryptography and, in particular,
serves as a basis for modern security proofs of Quantum Key
Distribution (QKD). We first explain this criterion and then describe
Hirota and Yuen's critique as well as the error in their argument.

\paragraph*{{\bf Secrecy in quantum cryptography.}} 
Realistic cryptographic keys (e.g., those obtained by QKD) are usually
not perfectly secret. Rather, their secrecy is quantified by a
parameter, $\eps \geq 0$, which bounds the maximum tolerated deviation
from an \emph{ideal key}, i.e., a key that is perfectly uniformly
distributed and independent of any information held by a potential
adversary. Formally, a key $S$ is said to be \emph{$\eps$-secret} if
the \emph{maximum advantage} for distinguishing $S$ from an ideal key
is at most $\eps$~\footnote{Consider a hypothetical experiment where a
  \emph{distinguisher} $D$, which has access to all information held
  by the adversary, is presented with one of two possible keys, $S$ or
  $\bar{S}$. The \emph{advantage} of $D$ for distinguishing $S$ from
  $\bar{S}$ is then defined as the difference between the probability
  that the distinguisher outputs $1$ in the two cases, i.e.,
  $\bigl|\Prob[{D(S) = 1}] -\Prob[{D(\bar{S}) = 1}]\bigr|$. Maximizing
  this over all possible distinguishers $D$ gives the \emph{maximum
    advantage} for distinguishing $S$ from $\bar{S}$.}.  This
definition guarantees that, in any application that is secure when
using an ideal key (such as one-time-pad encryption), one may also use
an $\eps$-secret key instead, with $\eps$ corresponding to the failure
probability caused by this replacement~\footnote{More precisely, when
  using an $\eps$-secret key $S$ in an application, the probability of
  any event (e.g., that an adversary can correctly guess an encrypted
  message) is upper bounded by $p + \eps$, where $p$ is the
  probability of the same event in an ideal scenario, where $S$ is
  replaced by an ideal key.}. The notion of $\eps$-secrecy enables
modular proofs of security, which is why one also refers to it as
\emph{universally composable (UC) secrecy}.

For the considerations below, we will assume that the key $S$ is a bit
string of length $\ell$ (a typical value is $\ell = 10^6$) and that
$\eps > 0$ is a small but strictly positive constant (for
concreteness, one may set $\eps = 10^{-20}$, which is achievable by
QKD and, at the same time, sufficient for all practical
purposes~\footnote{Since $\eps$ is an upper bound on the probability
  of a failure caused by the imperfection of the secret key (as
  explained above), it is generally sufficient to choose $\eps$
  smaller than the probability of a security breach due to other
  imperfections (e.g., a hardware problem leading to a key
  leakage).}).

Most modern quantum cryptographic security proofs, in order to
establish secrecy of a key $S$, rely on a mathematical criterion
introduced in~\cite{BHLMO05,RenKoe05}. The criterion is based on the
\emph{trace distance}~\footnote{For two density operators, $\rho$ and
  $\sigma$, the \emph{trace distance} is defined by $d(\rho, \sigma)
  := \frac{1}{2} \| \rho - \sigma\|_1$, where $\| \cdot \|_1$ denotes
  the trace norm.}, which we denote by $d(\cdot, \cdot)$, and demands
that
\begin{align} \tag{TD} \label{eq_TD}
  d\bigl(\rho_{S E}, \bar{\rho}_S \otimes \rho_E)  \leq
  \eps \ ,
\end{align}
where $\rho_{S E}$ denotes the joint state of the key $S$ and the
information $E$ held by the adversary, and $\bar{\rho}_S$ is a
completely mixed state (corresponding to a uniformly distributed $S$).
The use of this criterion is justified by the following
implication~\cite{BHLMO05,RenKoe05}
\begin{align} \label{eq_standard}
   (\text{\ref{eq_TD}})    \implies    (\text{UC secrecy}) \ .
\end{align}

\paragraph*{{\bf Recent scepticism.}} Hirota~\cite{Hirota12} and
Yuen~\cite{Yuen12b,Yuen12a,Yuen11c,Yuen11b,Yuen11a,Yuen10,Yuen09b,Yuen09a}
argue that the standard secrecy criterion~\eqref{eq_TD} does not
actually imply secrecy, i.e., that the above implication is wrong
(unless the parameter $\eps$ in~\eqref{eq_TD} is chosen exponentially
small in the key size). For concreteness, we refer in the following
specifically to the paper by Hirota~\cite{Hirota12}. We note, however,
that the argument is similar in spirit to Yuen's
reasoning~\cite{Yuen12b,Yuen12a,Yuen11c,Yuen11b,Yuen11a,Yuen10,Yuen09b,Yuen09a}
and, in fact, based on the latter.

The critique is built upon an alternative criterion that can be used
to establish the secrecy of a key $S$. The criterion demands that the
probability $P(S|E)$ that an adversary with knowledge $E$ can
correctly guess $S$ is small, i.e., \vspace{-2mm}
\begin{align} \label{eq_HY} \tag{HY} P(S|E) \sim 2^{-\ell}
\end{align}
where $\ell$ is the length of $S$. It is then argued that this
criterion is sufficient for secrecy, i.e.,
\begin{align} \label{eq_HirotaClaim}
  (\text{\ref{eq_HY}})    \implies    (\text{UC secrecy})  \ .
\end{align}
This implication is correct (if one takes the approximation
in~\eqref{eq_HY} to mean that the relative error between the left and
right hand side of $\sim$ is at
most~$\eps$)~\footnote{In~\cite{Hirota12}, this claim is formulated as
  part of a definition (Definition~3), which may have contributed to
  the confusion.}. Furthermore, by explicit examples (p.~5
of~\cite{Hirota12}), it is shown, again correctly, that~\footnote{To
  understand the difference between criteria~\eqref{eq_TD}
  and~\eqref{eq_HY}, one may consider the special case where $E$ is
  trivial, i.e., uncorrelated to $S$. In this case,
  criterion~\eqref{eq_TD} corresponds to the requirement that the
  probabilities of $S$ are \emph{on average} not much larger than
  $2^{-\ell}$ (the probabilities of a uniform distribution), whereas
  criterion~\eqref{eq_HY} demands that \emph{all} probabilities of $S$
  are (approximately) bounded by $2^{-\ell}$.}
\begin{align} \label{eq_stronger}
  (\text{\ref{eq_TD}}) \centernot\implies \text{(\ref{eq_HY})} \ .
\end{align}

Hirota now seems to argue that~\eqref{eq_HirotaClaim}
and~\eqref{eq_stronger} together imply that~\eqref{eq_standard} is
wrong. This conclusion is, however, logically wrong. It would only
hold if the implication in~\eqref{eq_HirotaClaim} went in the other
direction, i.e., if \eqref{eq_HY} was not only a \emph{sufficient},
but also a \emph{necessary} criterion for UC secrecy. But this is not
true, as one can convince oneself by a simple example~\footnote{Let
  $\eps = 10^{-20}$ and let $\bar{S}$ be an ideal (perfectly uniform
  and secret) key of length $\ell = 10^6$. Furthermore, let $S$ be a
  key that is identical to $\bar{S}$, except if $\bar{S}$ is equal to
  the zero string, $\bar{S} = \mathbf{0} = 00 \cdots 0$, in which case
  we set $S = \mathbf{1} = 11 \cdots 1$. Hence, by construction, the
  probability that $S$ deviates from the ideal key $\bar{S}$ is upper
  bounded by $2^{-\ell} \leq \eps$, i.e., $S$ is UC secret. However,
  the probability that $S = \mathbf{1}$ is twice as large as it should
  be for a uniform string. Hence, an adversary guessing $S=\mathbf{1}$
  would have a success probability of $2 \cdot 2^{-\ell}$, thus
  violating criterion~\eqref{eq_HY}.}.

We conclude by remarking that the claim of Hirota and Yuen, if it
would have been valid, would not only shake the foundations of quantum
cryptography, but have an equally drastic impact on classical
cryptography, where similar secrecy criteria are used~\footnote{For
  example, the definition of \emph{randomness extractors}|a concept
  widely used in cryptography|is based on a classical special case of
  criterion~\eqref{eq_TD} (with the trace distance replaced by its
  classical analogue, the \emph{variational} or \emph{statistical}
  \emph{distance}); see, e.g., \cite{Trevisan01}.}. However, as shown
here, their claim is false.


\begin{thebibliography}{1}
\bibitem{Hirota12} Osamu Hirota, Incompleteness and Limit of Quantum
  Key Distribution Theory, arXiv:1208.2106v2 (2012).

  \bibitem{Yuen12b} Horace P.\ Yuen, Unconditional Security In Quantum
    Key Distribution, arXiv:1205.5065 (2012).

  \bibitem{Yuen12a} Horace P.\ Yuen, Problems of Security Proofs and
    Fundamental Limit on Key Generation Rate in Quantum Key
    Distribution, arXiv:1205.3820 (2012).

  \bibitem{Yuen11c} Horace P.\ Yuen, Security Significance of the
    Trace Distance Criterion in Quantum Key Distribution,
    arXiv:1109.2675 (2011).

  \bibitem{Yuen11b} Horace P.\ Yuen, Fundamental And Practical
    Problems of QKD Security --- the Actual and the Perceived Situation,
    arXiv:1109.1066 (2011).

  \bibitem{Yuen11a} Horace P.\ Yuen, Problems of Existing
    Unconditional Security Proofs in Quantum Key Distribution,
    arXiv:1109.1051 (2011).

  \bibitem{Yuen10} Horace P.\ Yuen, Fundamental Quantitative Security
    In Quantum Key Distribution, Physical Review A 82, 062304 (2010).

  \bibitem{Yuen09b} Horace P.\ Yuen, Universality and The Criterion
    `d' in Quantum Key Generation, arXiv:0907.4694 (2009).

  \bibitem{Yuen09a} Horace P.\ Yuen, Key Generation: Foundations and a
    New Quantum Approach, IEEE Journal of Selected Topics in Quantum
    Electronics, vol.~15, pp.\ 1630--1645 (2009).

  \bibitem{BHLMO05} Michael Ben-Or, Michal Horodecki, Debbie W. Leung,
    Dominic Mayers, and Jonathan Oppenheim: The Universal Composable
    Security of Quantum Key Distribution, Proc.\ of TCC~2005, LNCS,
    Springer, vol.~3378, pp.\ 386--406 (2005).

  \bibitem{RenKoe05} Renato Renner and Robert K\"onig, Universally
    Composable Privacy Amplification Against Quantum Adversaries,
    Proc.\ of TCC~2005, LNCS, Springer, vol.~3378, pp.\ 407--425
    (2005).

  \bibitem{Trevisan01} Luca Trevisan, Extractors and Pseudorandom
    Generators, Journal of the ACM, vol.~48, pp.~860--879 (2001).

\end{thebibliography}
\end{document}